\documentclass[pdflatex,sn-mathphys-num]{sn-jnl}% Math and Physical Sciences Numbered Reference Style
\usepackage{graphicx}%
\usepackage{amsmath,amssymb,amsfonts}%
\usepackage{booktabs}%

\theoremstyle{thmstyleone}%
\theoremstyle{thmstyletwo}%

\theoremstyle{thmstylethree}%

\begin{document}
	
	\title{Commissioning of a fast fine-step electron-energy-scan system for electron-ion crossed-beams experiments}
	
	%%=============================================================%%
	%% GivenName	-> \fnm{Joergen W.}
	%% Particle	-> \spfx{van der} -> surname prefix
	%% FamilyName	-> \sur{Ploeg}
	%% Suffix	-> \sfx{IV}
	%% \author*[1,2]{\fnm{Joergen W.} \spfx{van der} \sur{Ploeg} 
		%%  \sfx{IV}}\email{iauthor@gmail.com}
	%%=============================================================%%
	
\author*[1,2]{\fnm{B. Michel} \sur{D{\"o}hring}}\email{michel.doehring@exp1.physik.uni-giessen.de}

\author[1]{\fnm{Alexander} \sur{Borovik Jr.}}\email{alexander.borovik@physik.uni-giessen.de}
%\equalcont{These authors contributed equally to this work.}

\author[1]{\fnm{Kurt} \sur{Huber}}\email{kurt.huber@strz.uni-giessen.de}
%\equalcont{These authors contributed equally to this work.}

\author[1]{\fnm{Alfred} \sur{M{\"u}ller}}\email{alfred.mueller@iamp.physik.uni-giessen.de}
%\equalcont{These authors contributed equally to this work.}

\author[1,2]{\fnm{Stefan} \sur{Schippers}}\email{stefan.schippers@uni-giessen.de}
%\equalcont{These authors contributed equally to this work.}

\affil*[1]{\orgdiv{I. Physikalisches Institut}, \orgname{Justus-Liebig-Universit{\"a}t Giessen}, \postcode{35392}, \city{Gie{\ss}en}, \country{Germany}}

\affil[2]{\orgdiv{Helmholtz Forschungsakademie Hessen f{\"u}r FAIR (HFHF)}, \orgname{GSI Helmholtzzentrum f{\"u}r Schwerionenforschung}, \postcode{64291}, \city{Darmstadt}, \country{Germany}}

	%%==================================%%
	%% Sample for unstructured abstract %%
	%%==================================%%
	
	\abstract{We report on the commissioning of a fast electron-energy scan system for measurements of electron-impact ionization cross sections. The Giessen crossed-beams experiment employs a high-power electron gun which has been developed over recent years and which permits wide variations of the beam parameters. A multi-electrode design enables the decoupling of the electron energy from the electron density. The newly implemented control system, which governs the various electrode potentials, is described together with the salient technical features.}

	
	%%\pacs[JEL Classification]{D8, H51}
	
	%%\pacs[MSC Classification]{35A01, 65L10, 65L12, 65L20, 65L70}
	
\maketitle
	
\section{Introduction}\label{sec:intro}

Electron-impact ionization (EII) of ions is a fundamental atomic collision process that governs the charge balance in plasmas. Accurate cross sections for single and multiple ionization of ions by electron impact are required for a reliable modeling and for a deeper understanding of such environments. To this end, the electron-ion crossed beams method has been employed in Giessen since four decades \cite[][and references therein]{Mueller2008a}. Recent measurements of EII cross sections addressed the modeling of plasma light sources for EUV lithography \cite{Borovik2013}, of fusion plasmas \cite{Schury2020}, and of kilonovae \cite{Doehring2025}, the ionization via resonances associated with $K$-shell excitation \cite{Ebinger2019}, as well as detailed benchmarking of quantum-theoretical cross section computations \cite{Liu2015,Jin2020a,Jin2024}.

Key to the successful exploitation of the Giessen electron-ion crossed beams setup has been the development of intense electron-beams \cite{Mueller1980,Mueller1980a,Achenbach1983,Achenbach1984,Stenke1995}. The electron gun, which has been our workhorse during the past decades, consisted of an electrically heated dispenser cathode, four sets of electrodes defining the potential in the electron-ion interaction region and providing beam-compression capabilities, and a water-cooled anode \cite{Becker1985}. At a maximum electron energy of 1000~eV, this gun delivered electron currents of up to 0.5~A, corresponding to particle densities of up to some $10^8$~cm$^{-3}$. The acceleration voltage was limited to 1000~V by the mechanical design of the gun, in particular by the distances between the various electrodes, which would have led to discharges if operated at higher voltages. 

In order to overcome the 1000-eV limit while keeping the possibility of producing very high electron currents, which is both particularly desirable for electron-impaction ionization studies with more highly charged ions with ionization energies of up to $\sim$1~keV, a new gun has been designed \cite{Shi2003a}, built, and put into operation \cite{Ebinger2017}. The new design, which is based on the design of the 1000-eV gun \cite{Becker1985}, has pushed the maximum electron energy and the maximum electron current to considerably higher values of 3500~eV and 0.9~A, respectively. In addition, there are --- as detailed below --- more electrodes as compared to the 1000-eV gun, allowing for a more flexible tailoring of the electron beam properties. The new 3500-eV gun has also served as a prototype for a 12.5-keV gun, which is currently used as an electron target in the heavy-ion storage ring CRYRING@ESR operated by the international Facility for Antiproton and Ion Research (FAIR) in Darmstadt, Germany \cite{Lestinsky2016}. 

The 3500-eV gun has already been successfully commissioned for the measurement of \emph{absolute} cross sections, where the animated-beams method \cite{Defrance1981,Brouillard1983,Mueller1985} is employed. In the present paper, we report on the implementation of a fast scanning technique, which allows one to measure \emph{relative} EII cross section as a function of energy with extremely low point-to-point uncertainties \cite{Mueller1988}. This facilitates the resolution of narrow resonance structures in electron-impact ionization cross sections as has been repeatedly demonstrated with our 1000-eV gun  \cite[e.g.][]{Mueller1988,Mueller1989,Mueller2014b,Ebinger2019}.

The present paper is organized as follows. Section~\ref{sec:exp} describes the general features of the Giessen electron-ion crossed-beams experiment. The 3500-eV electron gun and salient properties of the electron beam are described in Section~\ref{sec:egun} including the new energy-scan system.  First experimental results, which characterize the performance of the scan system, are presented in Section~\ref{sec:res}. Section~\ref{sec:sum} provides a summary and an outlook. Further details of the experimental equipment are collected in an appendix.

\section{The Giessen electron-ion crossed-beams setup}\label{sec:exp}
	
\begin{figure}[t]
	\centering\includegraphics[width=\textwidth]{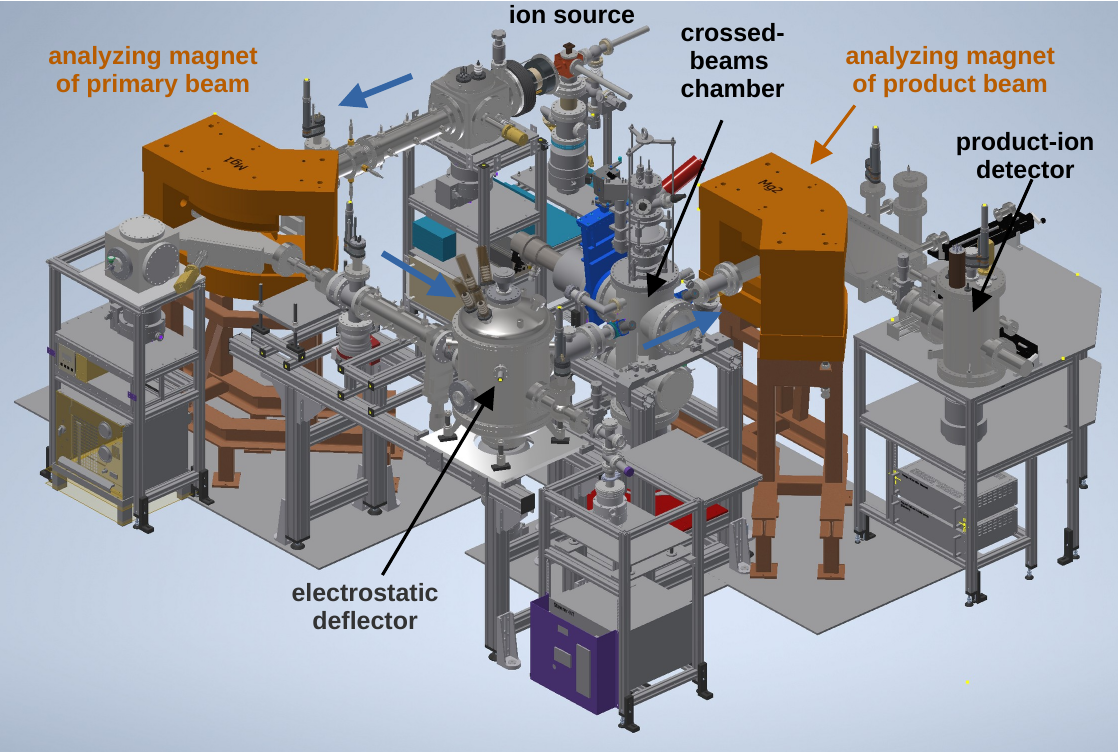}
	\caption{\label{fig:setup}CAD drawing of the Giessen electron-ion crossed-beams setup.  The blue arrows indicate the direction of the ion beam in the various sections of the setup. The footprint of the entire setup is about $4\times4$~m$^2$.}
		
	\end{figure}

The present Giessen electron–ion crossed-beams setup \cite{Teng2000a,Becker2004a,Jacobi2004a} has been in operation for more than two decades. Figure~\ref{fig:setup} provides an overview of the present arrangement, which has been relocated to a new laboratory very recently. Several types of (home-built) ion sources are available for the production of singly and multiply charged ions, such as a 10-GHz electron cyclotron resonance (ECR) source \cite{Trassl1997a}, a Penning discharge ion source \cite{Baumann1981}, and a Cs-sputter ion source for the production of negative ions \cite{Middleton1983}. The ion sources are operated on an electric potential $U_\mathrm{acc}$ of typically 12~kV. The ions are extracted towards the electrically grounded and evacuated beam pipe, whereby they acquire the kinetic energy $eqU_\mathrm{acc}$, where $e$ and $q$ denote the elementary charge and the ion charge state, respectively.  Ions of the desired mass-to-charge ratio $A/q$ are selected by passing the ion beam through a dipole magnet with appropriately adjusted magnetic field. Before entering the crossed-beams chamber, the ion beam is electrostatically deflected and then collimated by four-jaw slits with the apertures being adjustable in the range 0.1–-1.6~mm. The electrostatic deflection shortly before the electron-ion interaction region is intended to clean the beam from ions which have changed their charge state in collisions with residual-gas particles. 

In the crossed-beams chamber, the ion beam is crossed at 90$^\circ$ with the 6-cm wide ribbon-shaped electron beam of the above mentioned electron-gun, which is described in detail below. The process of $n$-fold EII of an ion $A^{q+}$, can be written as
\begin{equation}
\text{A}^{q+} + e^- \to \text{A}^{(q+n)+} + (n+1)e^-.
\end{equation}
After having left the electron-ion interaction region, the (few) A$^{(q+n)+}$ product ions are traveling with the same speed as the (still much more abundant) A$^{q+}$ primary ions. A second analyzing magnet is used to separate the various ion fractions in the product beam and to direct the product charge state of interest onto a channeltron-based single-particle detector \cite{Rinn1982}. The primary ions are collected in an appropriately positioned Faraday cup and the resulting ion current is monitored with a sensitive electrometer. The electron current, that reaches the electron gun's collector is measured in a similar manner. Absolute cross sections are derived by normalizing the measured detector count rate on the ion and electron currents while considering the geometrical beam overlap, which is determined implicitly by applying the animated-beams technique  \cite{Defrance1981,Brouillard1983,Mueller1985}  mentioned already above. The systematic uncertainties of the absolute cross sections are between 6.5\% and 10\%, and the statistical uncertainties are typically below 1\% at the 95\% confidence level, except near the ionization threshold where signal levels are low. 

The measurement of absolute cross sections  requires that the electron energy is constant during the measurement. Therefore, absolute cross section measurements can be only made with a relatively large temporal separation between the individual data points, such that additional point-by-point variations of, e.g., the geometric beam overlap can arise due to slow drifts of high-voltage power supplies. The  fast scanning technique \cite{Mueller1988}, the implementation of which is described below, aims at averaging out these slow variations by performing fast scans of the electron energy, where the dwell time per energy step is only a few milliseconds.

\subsection{Experiment Control and Data Acquisition}\label{sec:ecdaq}
	
The home-built electronics for experiment control and data acquisition is assembled in two 19-inch crates referred to as  \textit{control-routing} and  \textit{data-routing}, respectively. The crates have 20 slots each for Eurocard modules. Both units communicate via universal-serial-bus (USB) connections with a standard personal computer (PC). The latter operates under  the real-time operation system VxWorks. The hardware and associated software has been developed and expanded over more than 30 years \cite{VME-doku}. 

For controlling the various voltages of the  electron gun, parallel-to-serial output (PSO) cards located in the control-routing convert the digital control signals into a binary coded serial light signals produced by photodiodes (SFH756V) suitable for optical fibers of 980-\textmu{}m diameter (standard JIS C 6863). The use of optical fibers for signal transmissions ensures potential isolation and reliable signal transfer over fiber lengths of 1.5--20~m. On the receiving ends of the optical fibers, receiver diodes (SFH551V) feed the signals via home-developed FPGA (Field-Programmable Gate Arrays) firmware, which decodes the serial light signals, to 16-bit (LTC2752) or 18-bit (LTC2758) digital-analog-converters (DAC), which eventually provide  0--10~V analog programming voltages for the power supplies. In order to achieve a fine-grained control of the electron energy, the programming voltage of the electron gun's cathode power supply is provided by two stacked 18-bit DACs. The remaining gun electrodes are controlled more coarsely via 16-bit DACs. 

The data-routing system consists of modular interface cards (counters, clocks, Analog-Digital-Converters (ADC), etc.), which are tailored to the experimental signals to be fed into the data acquisition system. Each interface card is accompanied by a separate controller card, which ensures a synchronized transfer of the measured data to the PC system. The data are stored via TCP/IP on a hard disk of a Linux server.

\section{The 3500-eV electron gun} \label{sec:egun}

\begin{figure}[b]
	\centering\includegraphics[width=0.6\textwidth]{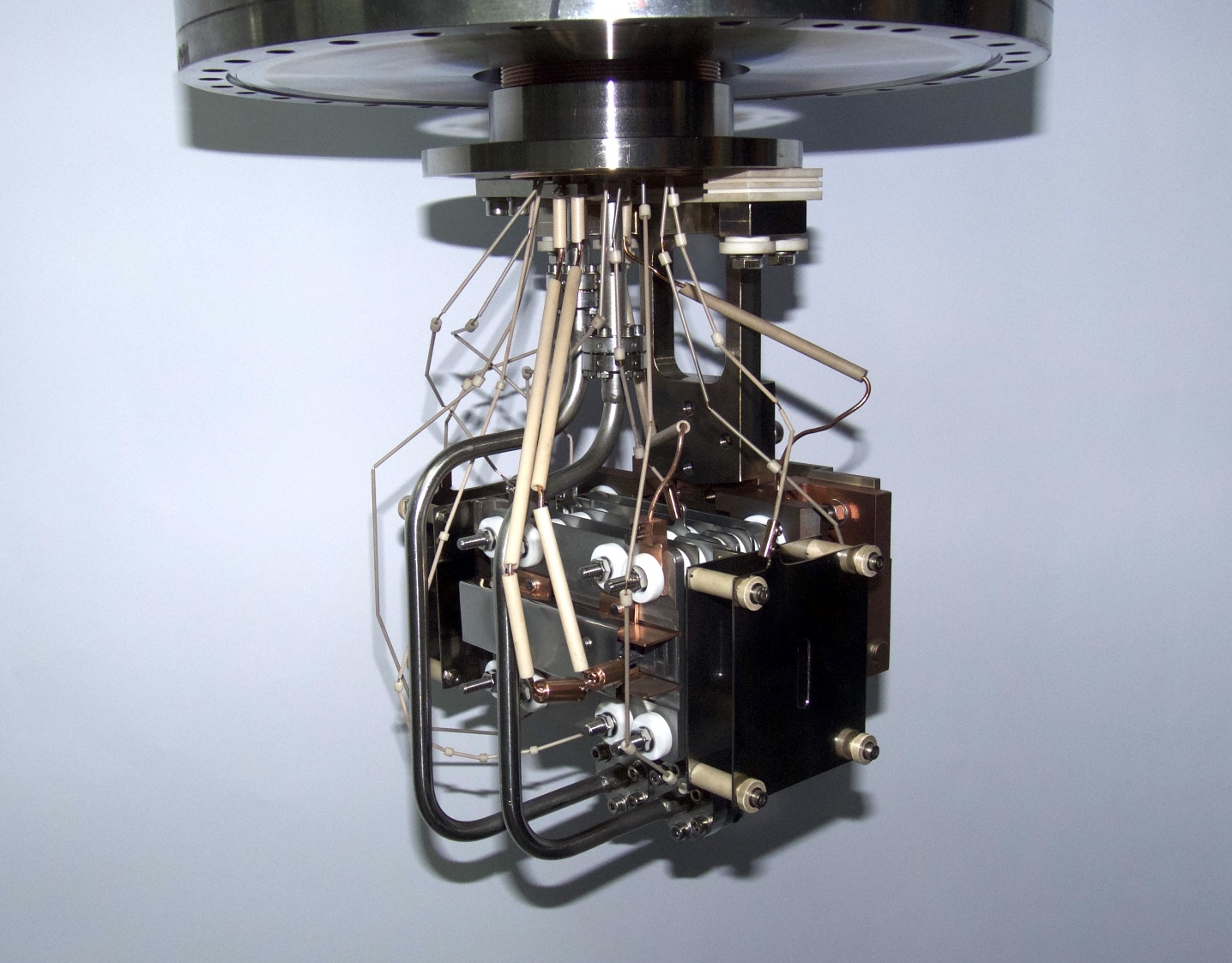}
	\caption{\label{fig:egun}Photograph of the 3500-eV electron gun before its insertion into the crossed-beams chamber (Fig.~\ref{fig:setup}). Each electrode is  individually electrically contacted. The stainless-steel tubes in the foreground belong to the cooling water circuit of the anode in the background. }
\end{figure}

\begin{figure}[t]
 \centering\includegraphics[width=0.9\textwidth]{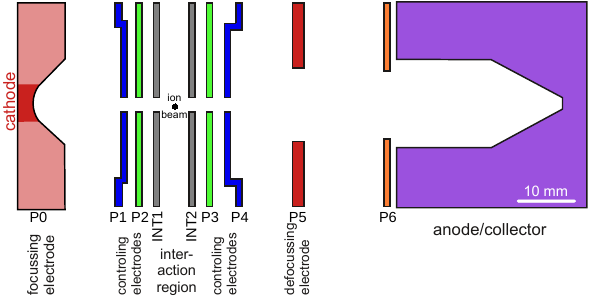}
 \caption{\label{fig:electrodes}Section along the electron beam direction providing an overview over the various electrodes of the 3500-eV electron gun. Electrodes of the same color are on the same electric potential. A detailed circuit diagram can be found in the appendix (Fig.~\ref{fig:circuit}).}
\end{figure}

A photograph of the 3500-eV gun is displayed in Fig.~\ref{fig:egun}, and Fig.~\ref{fig:electrodes} provides an overview over its various electrodes. Since its first commissioning in 2017 \cite{Ebinger2017}, the gun has undergone technical improvements, including a redesigned stainless-steel water-cooling system with CF flanges and silver gaskets, which reduced vibrations from the sometimes turbulent cooling water flow as compared to the original design.
 
The electrodes of the gun are designed such that the potential distribution is largely symmetric about the electron-ion interaction region.  The cathode is mounted and electrically connected to the focussing electrode P0. The curvatures of the cathode's emitting surface and of electrode P0 have been chosen for providing optimal beam focussing as predicted by electron optical simulations \cite{Shi2003a}. The potential difference $U_1$ between the cathode and electrode P1 determines the electron emission current $I_\mathrm{cath}$. The latter follows the relation $I_\mathrm{cath} = pU_1^{3/2}$ (with the perveance $p$) for space-charge limited electron emission. The electron energy is defined by the potential difference $U_e$ between the cathode and electrodes INT1 and INT2, which define the potential of the interaction region. When one introduces an effective perveance $p_\mathrm{eff}$, the relation between $I_\mathrm{cath}$ and $U_e$ can be approximately represented as (Fig.~\ref{fig:fecc}a)
\begin{equation}\label{eq:perv}
I_\mathrm{cath} = p_\mathrm{eff} U_e^{3/2}.
\end{equation}

Usually, INT1 and INT2 are on ground potential (0 V), but it is also possible to put them on a higher potential, e.g., for a\lq\lq{}voltage labelling\rq\rq\ of the product ions from the interaction region. The surrounding control electrodes P2 and P3, which are both on the same potential (different from INT1 and INT2), are responsible for shaping the electron beam in the interaction region. After having passed the interaction region, the electron beam is stepwise broadened with electrodes P4, P5 and  P6 before it is dumped into the grounded collector. 

In order to achieve an electron energy defined as accurately as possible and an electron energy spread as low as possible the power supplies must have excellent specifications with respect to reproducibility and to residual voltage fluctuations. For example, the cathode supply has a low ripple (peak-to-peak) of 20~meV plus $10^{-5}$ times the set voltage. Table \ref{tab:FUG-specs} in the appendix lists the specifications of all power supplies for the various electrodes. 

\subsection{Operation modes}

\begin{table}[t]
	\caption{Some tested operation modes of the electron gun. High-energy and high-current modes are designated as HE and HC, respectively. The potentials on the individual electrodes are given relative to the potential difference between the cathode and the interaction region. In this unit, the potential of the cathode and of electrode P5 is $-1$. The potential of P6 is adjusted during operation to values between 0.1 and 1.0 as required for a minimization of loss currents. 'INT' refers both to electrode INT1 and INT2. The column labeled 'Col.' gives the potential of the collector. The listed effective perveances (Eq.~\ref{eq:perv}) were determined experimentally (Fig.~\ref{fig:fecc}a).}
	\label{tab:modes}
	\begin{tabular}{@{}lccccccc@{}}
		\toprule
		\multicolumn{1}{c}{Mode}  & \multicolumn{6}{c}{Electrode}  & Effective Perveance  \\ 
	&	\multicolumn{1}{l}{P1} &
		\multicolumn{1}{l}{P2} &
		\multicolumn{1}{l}{INT} &
		\multicolumn{1}{l}{P3} &
		\multicolumn{1}{l}{P4} &
		\multicolumn{1}{l}{Col.} & 
        [$10^{-6}$~AV$^{-3/2}$] \\
        \midrule
		HE10  & 0.10 & 0.10 & 0 & 0.10 & 0.10 &  0 &  \phantom{7}4.44 \\
		HE15  & 0.15 & 0.15 & 0 & 0.15 & 0.15 & -1 &  \phantom{7}4.79 \\
		HC6   & 6.00 & 0.25 & 0 & 0.25 & 6.00 &  0 & 74.41  \\ \bottomrule
	\end{tabular}
\end{table}

\begin{figure}[b]
	\centering\includegraphics[width=\textwidth]{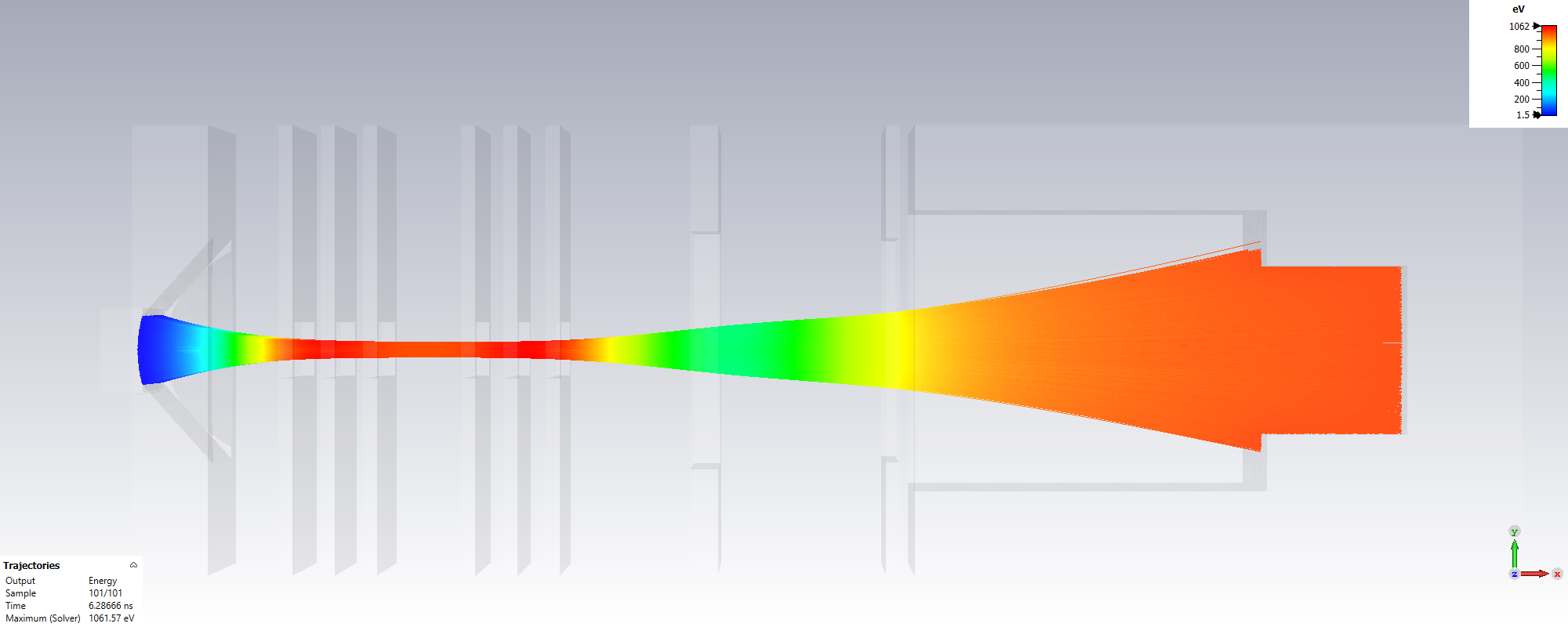}
	\centering\includegraphics[width=\textwidth]{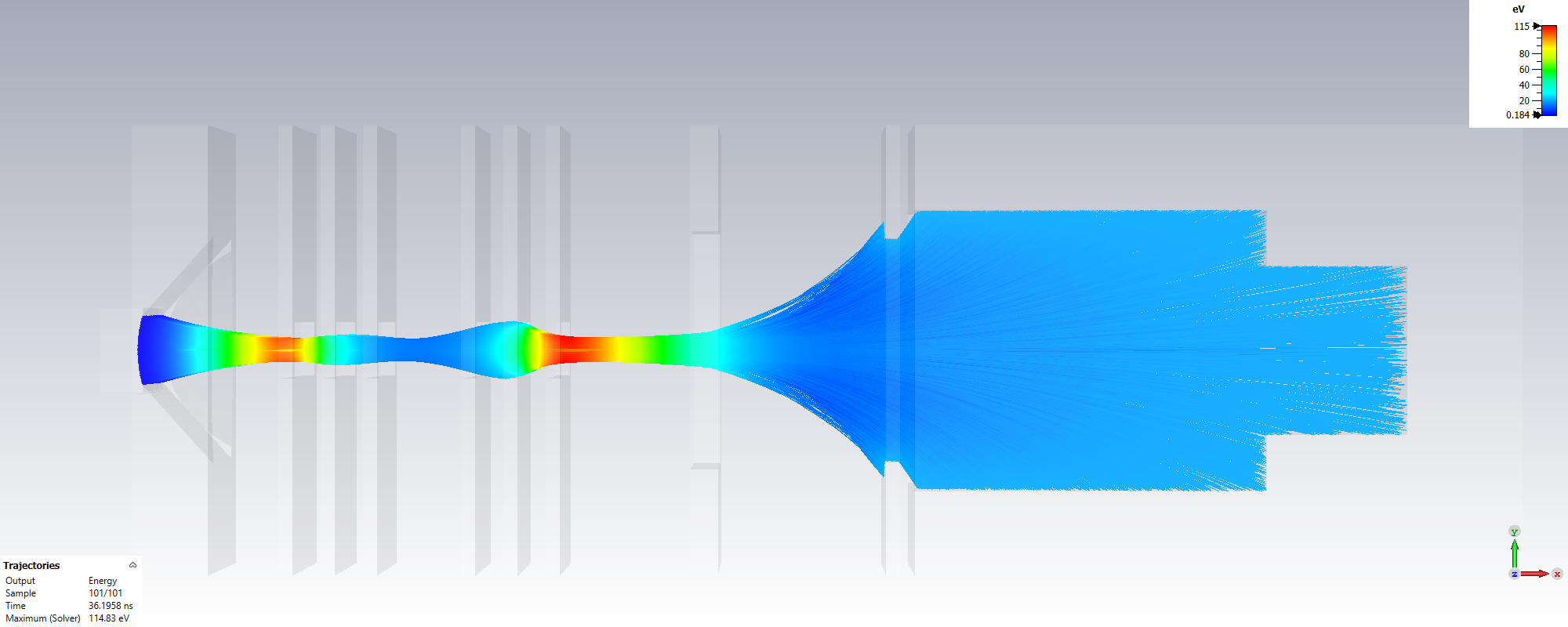}
	\caption{\label{fig:simul}Simulated electron trajectories in the 3500-eV electron gun. Top: Mode HE10 with $U_e$~=~1000~V and $I_\mathrm{cath} = 147.351$~mA. Bottom: Mode HC6 with $U_e = 20$~V and $I_\mathrm{cath} = 7.14$~mA. The color encodes the electron energy. The simulations were carried out with the software CST Studio Suite \cite{CST,Spachmann2006}.}
	
\end{figure}

The combination of in total ten electrodes decouples electron energy from electron emission and allows for adjustments of the power loads on the gun electrodes. For setting the voltages of the various electrodes  during the 2017 commissioning \cite{Ebinger2017} we have been guided by electron-optical simulations \cite{Shi2003a} and found good agreement between the predicted and experimentally realized electron gun performance. However, during the subsequent use of the 3500-eV gun in electron-ion collision experiments, we have observed that some combinations of electrode potentials lead to significant deflections of the ion beam when the electron gun was moved during animated-beams measurements and which compromised the absolute cross section measurements. Table~\ref{tab:modes} lists some voltage settings (which we refer to as \lq\lq{}operation modes\rq\rq), which provide conditions compatible with the requirements of cross-section measurements. The high-current mode HC6 is particularly suited for measurements at comparatively low electron energies up to 60~eV. At higher energies high-energy (HE) modes should be used. Figure~\ref{fig:simul} shows results of electron-optical simulations. Accordingly the electron beam trajectories are drastically different in the high-energy and high-current modes. 

\subsection{Loss-current monitoring}

\begin{figure}[t]
	\centering\includegraphics[width=0.95\textwidth]{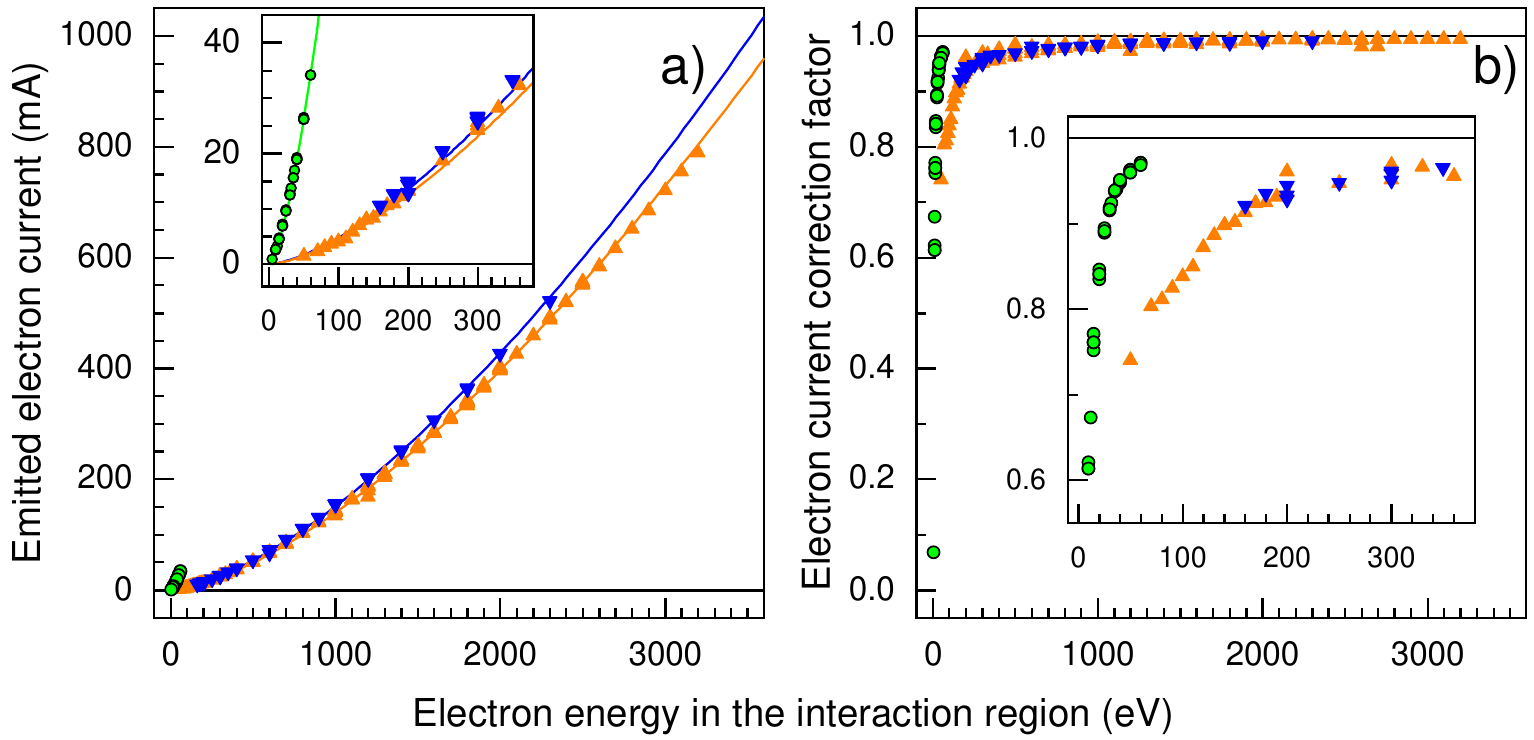}
	\caption{\label{fig:fecc} Measured (panel a) emitted electron currents $I_\mathrm{cath}$ and (panel b) electron current correction factors  $f_\mathrm{ecc}$ (Eq.~\ref{eq:efcc}) as functions of electron energy for the operation modes from Tab.~\ref{tab:modes}, i.e., HE10 (orange up-triangles), HE15 (blue down-triangles), and HC6 (green circles). The insets zoom into the low-energy regions. The full lines in the left panels represent fits of Eq.~\ref{eq:perv} to the data points, which result in the tabulated values for the effective perveance in Tab.~\ref{tab:modes}. }
	\label{pic:Egun-ECC-transmission}
\end{figure}

\begin{figure}[t]
	\centering\includegraphics[width=0.8\textwidth]{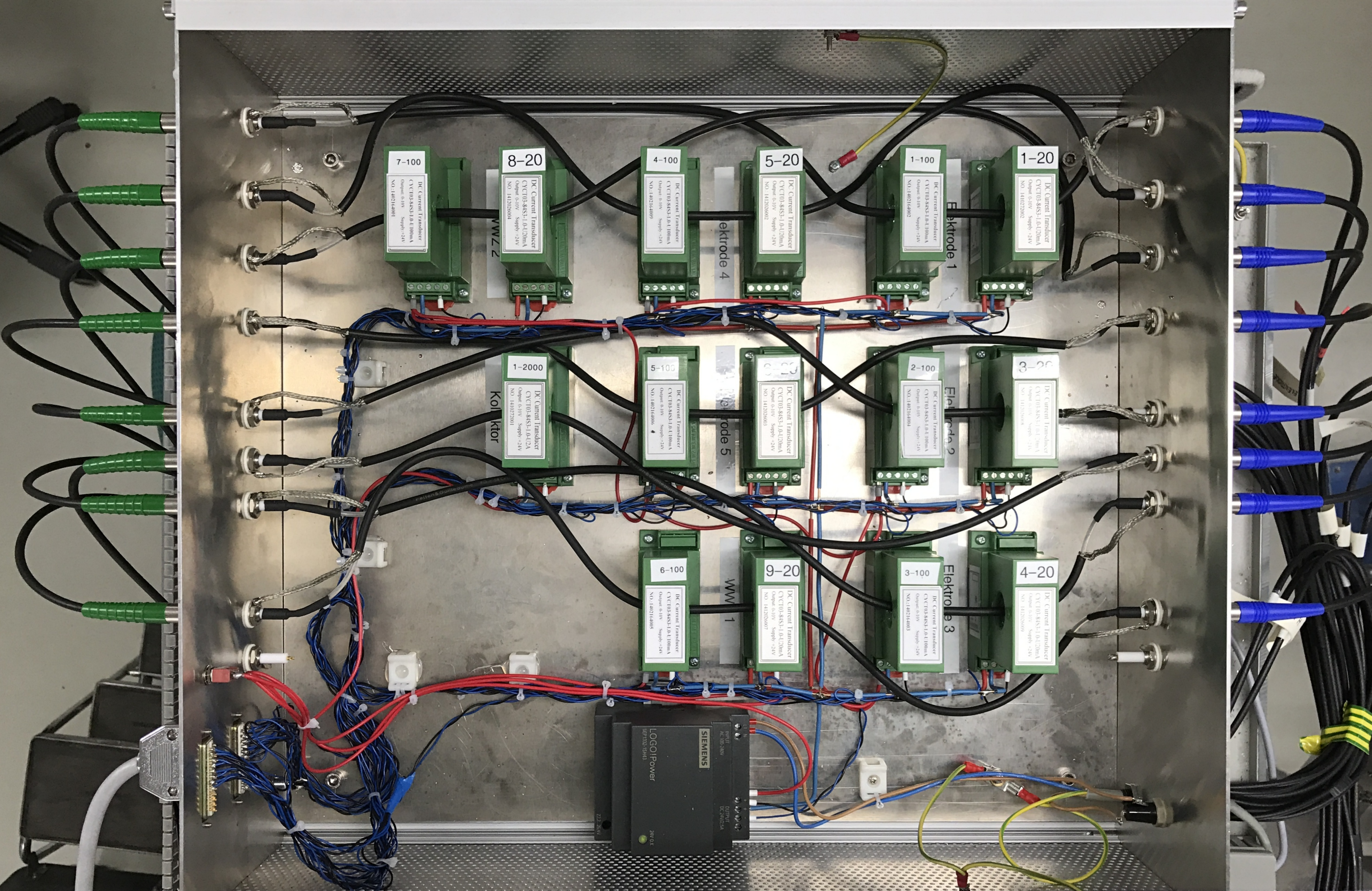}
	\caption{\label{fig:sensbox}View into the metal box (upper lid removed), that contains the hall-effect sensors for monitoring the loss currents on the various electrodes of the 3500-eV electron gun. The SHV plugs attached to the outsides of the box are colored green and blue for the connections to the inputs and outputs, respectively, of the sensors. There are also currently unused Hall probes with 100-mA range in series with some of the 20-mA Hall probes.}
	
\end{figure}

The flexibility of our electron gun permits operation over a broad energy range but also increases the risk of loss currents by electrons at mid and high energies. Excessive loss currents, which are partly dissipated in heat, lead to severe damage of the electrodes and of the ceramic spacers that are used to electrically isolate the electrodes from one another. This was observed in the early stages of the commissioning of the 3500-eV gun. The risk of loss-current induced damage was mitigated by changing the electrode material from copper to molybdenum (except for the collector) and by installing an emergency shutdown system, which is based on  monitoring the electric currents through the connecting wires of the electrodes by contact-free Hall-effect sensors (ChenYang type CYCT03-xnS3 \cite{ChenYang-sens}, Fig.~\ref{fig:sensbox}) with current ranges of 0--2000~mA for the collector and of 0--20~mA for the other electrodes.  The 0--10~V output voltages of the Hall-effect sensors are noise filtered and subsequently digitized using ADC cards equipped with a 12-bit  ADC chip (Microchip MCP3208B).  The emergency shutdown system compares the measured loss currents with adjustable hardware thresholds and in case of current overruns automatically switches off the gun within half a second to prevent damage. 

The software control system has additional stop and warn levels which are typically lower than the hardware thresholds and which are used to control the speed of voltage changes during electron-energy scans to prevent too drastic changes, which would cause high loss currents. This is particularly important for scans which are supposed to run over several days or even weeks as has been done in the past with the 1000-eV~gun (see, e.g., \cite{Mueller2014b}). 

The data acquisition system records the continuously monitored loss currents during cross section measurements. These data can be used for estimating the electron current $I_e$ in the electron-ion interaction region, which is the decisive quantity for the determination of absolute cross sections. It is calculated as 
\begin{equation}
    I_e = I_\mathrm{cath} - I_\mathrm{P1} - I_\mathrm{P2} - I_\mathrm{INT1}
\end{equation}
where $I_\mathrm{cath}$ is the electron current emitted from the cathode and $I_\mathrm{P1}$, $I_\mathrm{P2}$, and $I_\mathrm{INT1}$ are the loss currents measured at the electrodes P1, P2 and INT1 (Fig.~\ref{fig:electrodes}). At low electron energies, where beam focussing is poor, the electron current correction factor
\begin{equation}\label{eq:efcc}
    f_\mathrm{ecc} = \frac{I_e}{I_\mathrm{cath}} = 1 - \frac{I_\mathrm{P1} - I_\mathrm{P2} - I_\mathrm{INT1}}{I_\mathrm{cath}}
\end{equation}
is considerably smaller than 1 (depending on operation mode), but rapidly approaches unity with increasing electron energy (Fig.~\ref{fig:fecc}b). For high-energy modes it is basically constant above an electron energy of 800~eV.

\subsection{Energy-scan system}\label{sec:scan}

\begin{figure}[htp]
	\centering\includegraphics[width=1\textwidth]{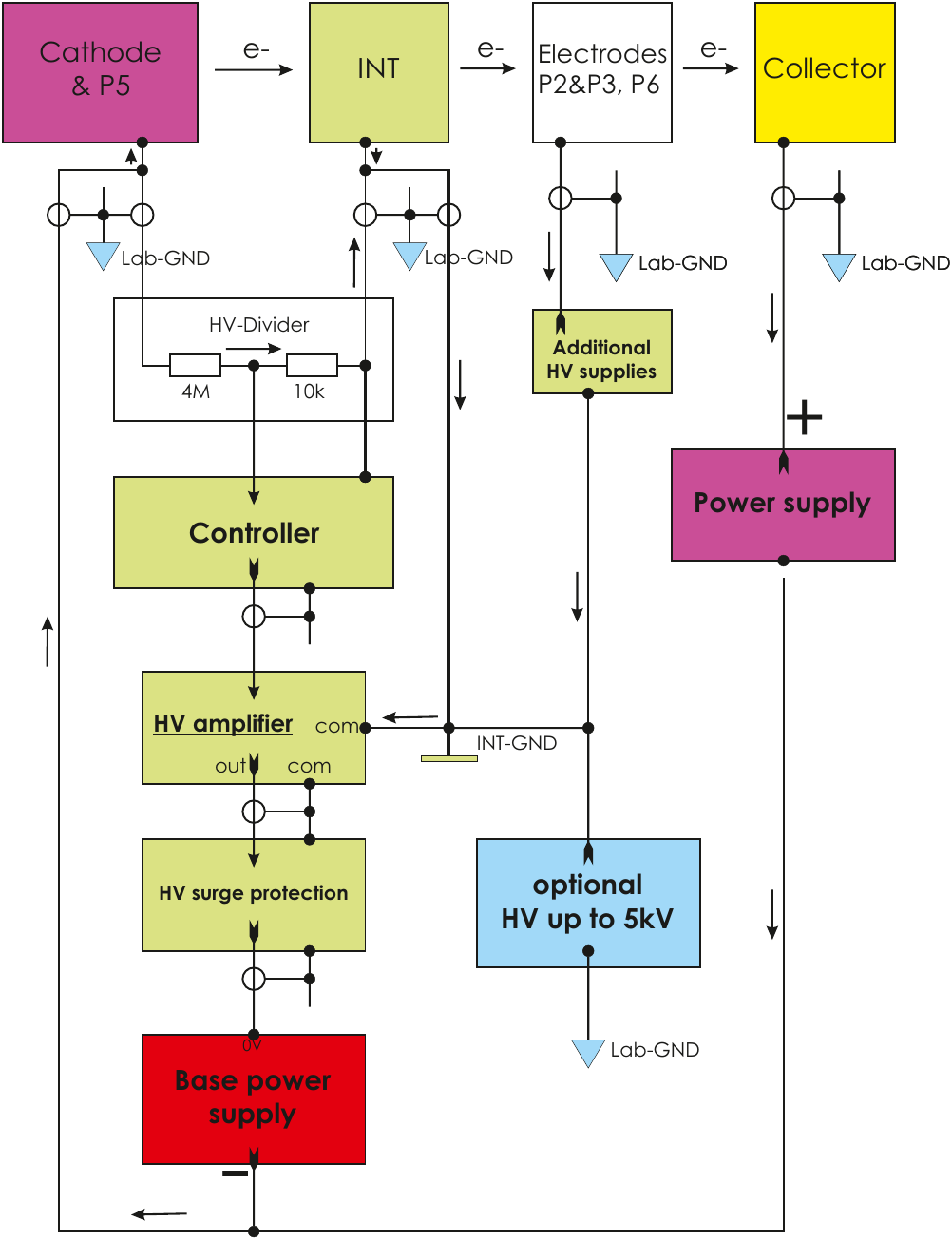}
	\caption{\label{fig:loop}Overview of voltage and current flow for the flexible gun. To keep the diagram simple, the high-voltage control for the electrode pair P1\&P4 is not shown. It is a copy of the circuit for the pair cathode\&P5 but uses a $1:200$-voltage divider instead of the $1:400$-divider (see Fig.~\ref{fig:filter} in the appendix for details). Since the block diagrams for the electrode pair P2\&P3 and for electrode P6 would be identical, only one is provided for the sake of simplicity of the drawing. The arrows indicate negative currents. All modules are drawn without chassis ground.}
\end{figure}

An energy scan through a preselected range of energy steps requires a synchronous control of all electrode power supplies such that the voltage ratios of a given operation mode (Tab.~\ref{tab:modes}) are (approximately) maintained at each energy step. In principle, this can be achieved with the above mentioned DACs (Sec.~\ref{sec:ecdaq}). However, the accuracy of the power supplies is too low for assuring a sufficiently accurate reproducibility of the electron energy when repeatedly ramping through the preselected scanning range. A measurement consists of typically several hundred repetitions until a satisfying level of statistical uncertainty is reached.   

In order to improve on the repeatability of the electron energy, we have established a  feedback loop where the voltages  of the cathode and of electrode P1, which determines the electron emission current, are measured with voltage dividers and where the control voltages of the fast-high voltage amplifiers are swiftly adjusted such that the deviation between the desired voltages and the read voltages is minimized on a sub-millisecond timescale. The accuracy of the voltages on the focussing and defocussing electrodes is less critical. Therefore, the corresponding power supplies are directly controlled by their DACs without feedback loops.      

The voltage divider ratios are 1:400 for the cathode and 1:200 for electrode P1. Their high-stability resistors have a temperature stability of 1~ppm per degree Celsius. Their ratings are accurate to within 0.1\%. Since the two high-voltage amplifiers  (Kepco BOP 100-1M) for the cathode and for electrode P1 have an output range of only $\pm100$~V they are each electrically connected in series with a base power supply as shown in Fig.~\ref{fig:loop} for the cathode. A controller box generates the control voltage for the high-voltage divider from the comparison of the voltage divider output with the set value. The latter is supplied by adding the outputs of two 18-bit DACs, where the first one covers a voltage range of 0--4000~eV with a coarse resolution of 15.25885~mV and the second one covers a voltage range of 0--400~eV with a fine resolution of 1.525885~mV. A third DAC is used to set the output voltage of the base power supply to approximately the center of the scan range. As long as the voltage span of a scan is less than 200~eV the third DAC is kept at a fixed value. For larger scan ranges the third DAC is adjusted from time to time as required for covering the entire preselected scan range.  

To reduce electrical noise, identical cable types and lengths have been used for all critical connections. In addition, a passive filter box (Fig.~\ref{fig:filter}) damps the remaining electric noise without compromising the sub-millisecond time constants of the voltage feedback loops. Immediately after power-up, the HV control units require a $\sim$2-h warm-up to reach stable conditions as has been determined by a 24-h monitoring of the control voltages. We also found, that changes in the ambient condition, e.g.\ by opening and closing of windows or doors, can induce millivolt-level perturbations. In order to minimize such effects, the control units have been mounted in 19-inch racks with temperature-controlled ventilation.

The specified voltage slew rate of the high voltage amplifiers is 11~V/\textmu{}s, i.e., the amplifier can accommodate typical voltage changes of about 20~meV in much less than a microsecond. In practice, the time constant for reaching stable conditions after a voltage change is much longer, since it is determined by various capacities in the electric circuit. Therefore, in our energy scans we allow for a waiting time before starting the data acquisition after setting a new voltage. A waiting time of 0.3~ms has been found to be sufficient for reaching stable conditions.  The dwell time per energy step is typically a few milliseconds.

\section{Performance of the energy-scan system}\label{sec:res}

\begin{figure}[b]
	\centering\includegraphics[width=0.95\textwidth]{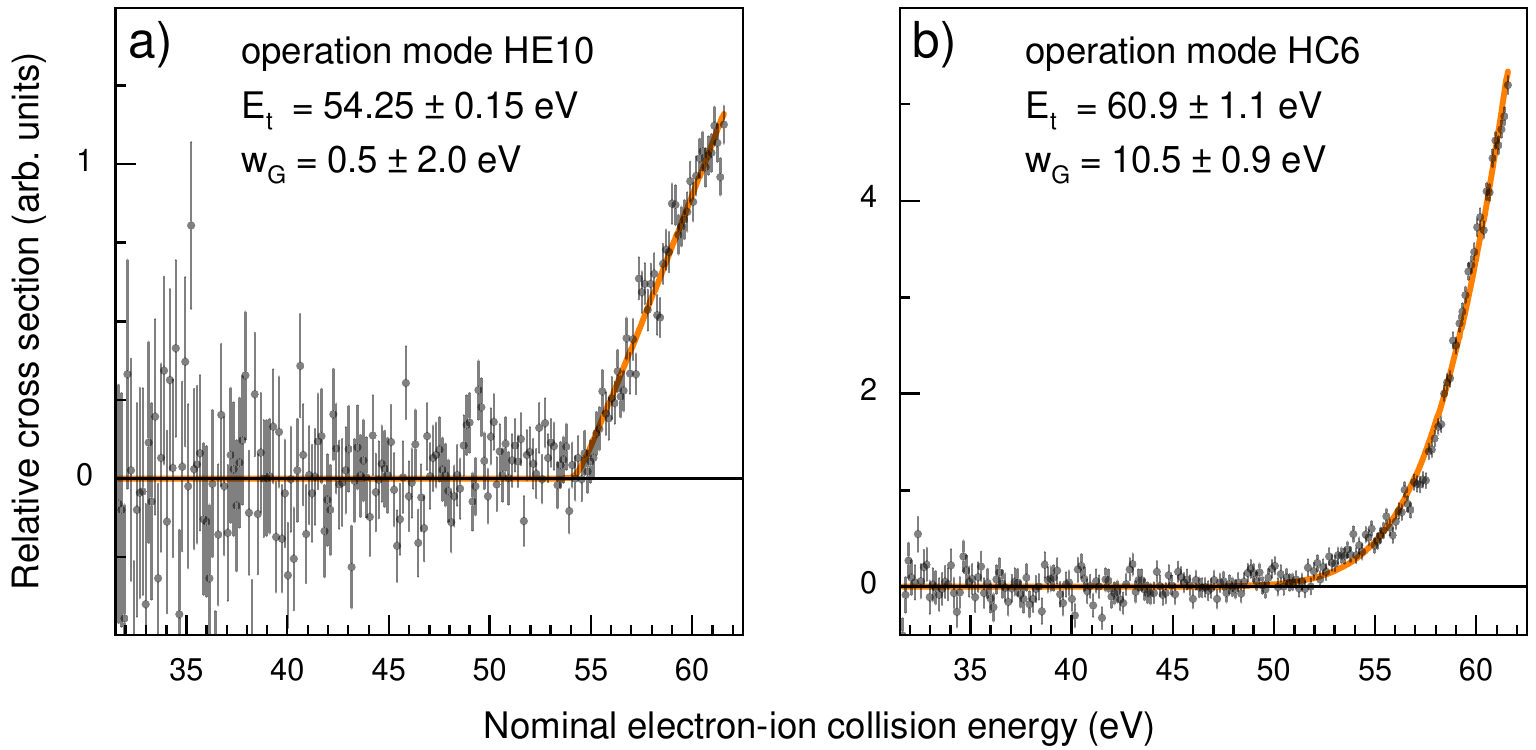}
	\caption{\label{fig:Hethres}Relative cross sections (symbols) for electron-impact ionization of He$^{+}$ as functions of nominal electron-ion collision energy measured with operation modes (Tab.~\ref{tab:modes}) HE10 (panel a) and HC6 (panel b). The total measurement times per data point were 3.6~s and 2~s, respectively. Each spectrum consists of 201 data points. The full lines are fits of Eq.~\ref{eq:WG}, i.e., the Wannier threshold law (Eq.~\ref{eq:Wannier}) convolved with a normalized Gaussian, to the experimental data. The threshold energies $E_t$ and Gaussian FWHM $w_G$ obtained from the fits are provided for each fit along with their fit uncertainties. The relative cross section scales in both panels are independent from each other. }
\end{figure}

Figure~\ref{fig:Hethres} shows the relative cross section for electron impact ionization of hydrogenlike He$^+$ ions in the vicinity of the ionization threshold, which occurs at 54.418~eV \cite{Kramida2024}. Such threshold scans can be used for calibrating the experimental energy scale. The experimental uncertainty of the nominal energy scale is rather large. In addition to the only 0.1\% accuracy of the resistors in the voltage dividers, there are also (unknown) contact potentials and uncertainties due to the electron beam's space charge. 

\begin{figure}[b]
	\centering\includegraphics[width=0.7\textwidth]{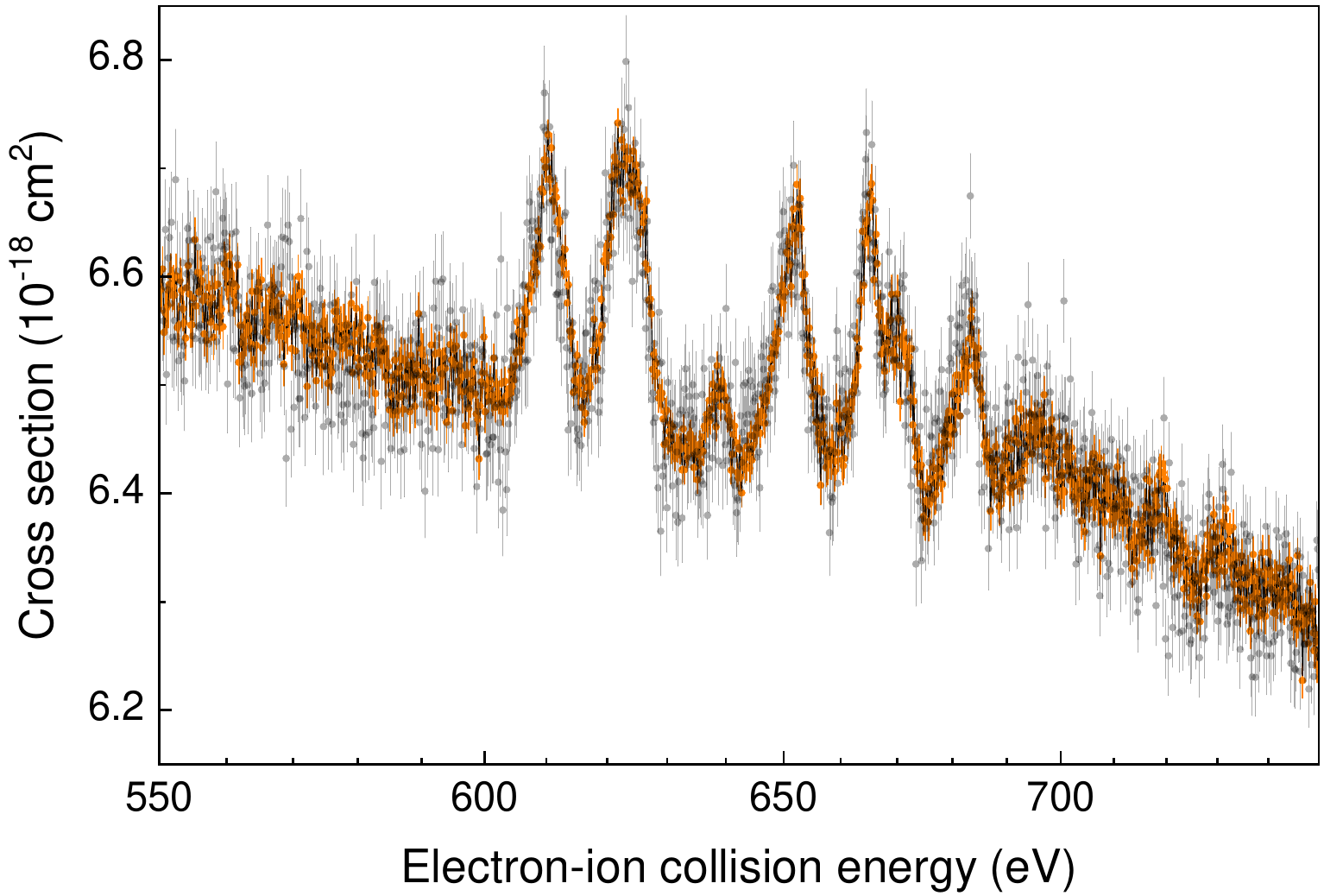}
	\caption{\label{fig:Xe9} Experimental cross section for electron-impact single ionization of Xe$^{9+}$ ions. The results from the new 3500-eV gun (transparent black symbols) are compared with earlier results from the old 1000-eV gun (orange symbols) \cite{Borovik2015}. With the 3500-eV gun, the operation mode HE10 was used. The ion beam was collimated to a size of  $1.4\times 1.4$~mm$^2$. The ion  and electron currents in the interaction region were $I_{\mathrm{ion}}=5.2$~nA and $I_{\mathrm{e}}\approx 82.2$~mA. The total measurement time per data point was 134~s.} 
\end{figure}

A commonly assumed behavior for the threshold cross section $\sigma_\mathrm{thres}(E)$ is Wannier's law \cite{Wannier1953}. Accordingly,
\begin{equation}\label{eq:Wannier}
\sigma_\mathrm{thres}(E)= S(E-E_\mathrm{thres})^{\alpha},
\end{equation} 
where $S$ is an overall scaling factor, $E_\mathrm{thres}$ denotes the threshold energy, and the constant $\alpha$ depends on the ion charge state $q$. The full curves in Fig.~\ref{fig:Hethres} were obtained by fitting Eq.~\ref{eq:Wannier} with $\alpha=1.056$ for $q=1$. In order to account for the electron energy spread, Eq.~\ref{eq:Wannier} was convolved with a normalized Gaussian with its full width at half maximum (FWHM) denoted as $w_G$. The convolved cross section, which was actually used as a fit function, reads
\begin{eqnarray}
\lefteqn{\sigma_\mathrm{thres}^{(G)}(E) = \frac{S}{\sqrt{\pi}} \left(\frac{w_G}{\sqrt{4\ln2}}\right)^\alpha\times} \label{eq:WG}\\ & & \left[\frac{1}{2}\,\Gamma\left(\frac{1}{2}+\frac{\alpha}{2}\right)\, _{1\!}F_1\left(-\frac{\alpha}{2},\frac{1}{2}, -\varepsilon^2 \right)+
   \varepsilon\,\Gamma\left(1+\frac{\alpha}{2}\right)\, _{1\!}F_1\left(\frac{1}{2}-\frac{\alpha}{2},\frac{3}{2}, -\varepsilon^2 \right) \right],\nonumber
\end{eqnarray}
where $\Gamma$ and $_{1\!}F_1$ are the gamma function and the confluent hypergeometric function, respectively, and the dimensionless quantity
\begin{equation}
\varepsilon = \sqrt{4\ln2}\,\frac{E-E_\mathrm{thres}}{w_G}
\end{equation}
is a function of the electron-ion collision energy $E$. 

In the high-energy mode, the statistical accuracy is too low for obtaining a reasonably small fit uncertainty for the experimental energy spread (Fig.~\ref{fig:Hethres}). In any case, the energy spread is much lower in the high-energy mode than in the high-current mode. There is also a significant discrepancy in the fitted threshold energies. In the high-energy mode, the fitted threshold value agrees within its uncertainty with the literature value. In the high-current mode, the threshold appears at a 6.5~eV higher nominal energy. This indicates that the space-charge potential of the electron beam is significantly larger in the high-current mode than in the high-energy mode. In both cases, the He$^+$ ionization cross section provides a valuable calibration point with a calibration uncertainty of the order of 0.2~eV (1.0~eV) in the HE10 (HC6) mode. In case of the high-energy mode, the calibration uncertainty can certainly be improved by accumulating better statistics. Further calibration points can be obtained in a similar manner from threshold cross-section measurements with other ions.

At higher energies, Fig.~\ref{fig:Xe9} compares cross-section scans for electron-impact single ionization of Xe$^{9+}$ which were measured with our old 1000-eV electron gun \cite{Borovik2015} and with our new 3500-eV gun. In the energy range 600--700~eV, the cross section exhibits resonance features which are associated with the excitation of a $3d$ electron to a higher shell and  simultaneous capture of the initially free electron. Together with a subsequent two-step Auger decay, which then leads to net single ionization, the process is termed resonant-excitation--double-autoionization (REDA). Apart from differences in the statistical uncertainties both resonance spectra agree excellently with one another. This testifies that our newly implemented electron-energy scan system performs as desired. In particular, the electron energy spreads of the old gun and the new gun are about the same as is evident from the close similarity of the measured resonance widths and heights. Since one experimental resonance peak most likely comprises several atomic resonances one cannot derive the experimental energy from the measured cross sections without a detailed theoretical description of the REDA resonances.  
	 
\section{Summary and Outlook}\label{sec:sum}
	
Building on our experience with our old 1000-eV electron gun, we have successfully implemented and commissioned a fast electron energy scanning system for our new 3500-eV electron gun. This gun has more electrodes than the previous one and can, therefore be operated more flexibly.  For the time being we have established high-energy and high-current operation modes, where the latter ones can be used to prepare intense low-energy electron beams. Our first measurements show, that this leads to strong space-charge effects which compromise the experimental energy resolution. When using high-energy modes the conditions of the 1000-eV electron gun are recovered.

In the future we will explore further possibilities. An interesting aspect of the more flexible electrode layout is the originally envisaged \cite{Shi2003a} possibility to counteract the space charge potential of the electron beam. This holds the promise of providing a much higher energy resolution, which would be particularly beneficial for the study of ionization resonances.

	\backmatter
	
	\bmhead{Acknowledgements}
	The authors thank Benjamin Ebinger and Florian Gocht for their assistance in the early stages of the present work and during some of the pertaining measurements, respectively. 
This research has been funded in part by the German Federal Ministry for Research, Technology and Space (BMFTR) within the ErUM-Pro funding scheme under contracts 05P21RGFA1 and 05P24RG2.

	\bmhead{Data availability}
	The datasets generated and analysed during the current study are available from the corresponding author on reasonable request.

	\newpage
	\onecolumn
	%%===================================================%%
	%% Appendix                                          %%
	%%                                                   %%
	%%===================================================%%	
		\begin{appendices}

\section{Technical specifications}\label{sec:technical}

\begin{table}[htp]
\caption{\label{tab:FUG-specs}Specifications of HV power supplies (according to the manufacturers) for the electron gun. Values are given as "less-than" and pp is short for peak-to-peak. For a detailed circuit diagram see Fig:~\ref{fig:circuit}. The manufacturer of the BOP high-voltage amplifiers is Kepco. All other listed power supplies were manufactured by FUG. This information is given for
technical completeness only and does not imply any evaluation, endorsement, or preference over similar products.}
\begin{tabular}{p{3cm}lll}
\toprule
 &  set point  &                 &                 \\ 
 &  resolution & residual ripple & reproducibility \\ 
 \midrule
\underline{Cathode/P0/P5}\\ HCP 140-3500\\ (3550~V / 40~mA)        & $\pm5\cdot10^{-4}$ & $1\cdot10^{-5}$~pp $+$ 20~mV$_\mathrm{pp}$    & $\pm1\cdot10^{-5}$ \\[1ex] 
\textit{Fine setting:}\\ BOP 100-1M\\ ($\pm100$~V / 1~A)           & 1.53~mV            & $<10$  to 30~mV$_\mathrm{pp}$                 & $\pm1\cdot10^{-4}-5\cdot10^{-5}$\\[2ex] 
\underline{Cathode Heating}\\ NTN 350-20\\ (20~V / 15~A)           & $\pm1\cdot10^{-4}$& $3\cdot10^{-4}$~pp                            & $\pm5\cdot10^{-4}$ \\[2ex]  
\underline{Electrodes P1/P4}\\ MCP 140-2000\\ (2000~V / 60~mA)     & $\pm5\cdot10^{-4}$ & $\pm5\cdot10^{-5}$~pp $+$ 50~mV$_\mathrm{pp}$ & $\pm1\cdot10^{-4}$ \\[1ex]
\textit{Fine setting:}\\ BOP 100-1M\\ (100~V / 1~A)                & 1.53~mV            & $<10$  to 30~mV$_\mathrm{pp}$                 & $\pm1\cdot10^{-4}-5\cdot10^{-5}$\\[2ex]
\underline{Electrodes P2/P3}\\ Amplifier 609E-6\\ (4000~V / 20~mA) & $1\cdot10^{-3}$    & 80~mV rms                                     & N/A       \\[2ex]
\underline{Electrode P6}\\ HCN 140-6500\\ (6500~V / 20~mA)         & $\pm5\cdot10^{-4}$ & $\pm5\cdot10^{-5}$~pp $+$ 50~mV$_\mathrm{pp}$ & $\pm5\cdot10^{-4}$ \\[2ex]
\underline{Collector}\\ HCN 4200-3500\\ (3500~V / 1200~mA)         & $\pm5\cdot10^{-4}$ & $\pm2\cdot10^{-4}$~pp $+$ 200~mV$_\mathrm{pp}$& $\pm5\cdot10^{-4}$ \\
\bottomrule                                                             
\end{tabular}
\end{table}

\begin{figure}[htp]
	\centering\includegraphics[angle=90,width=\textwidth]{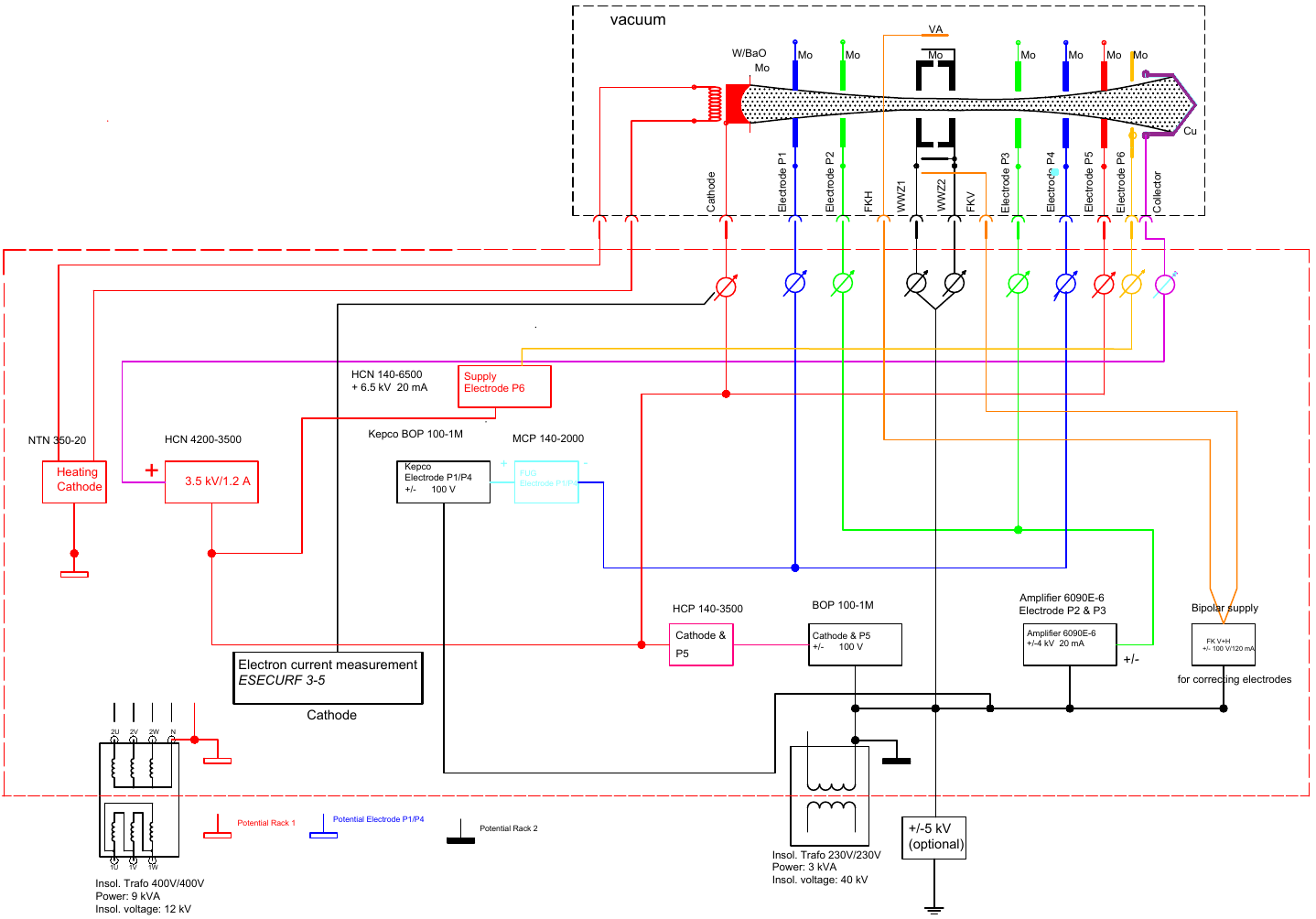}
	\caption{\label{fig:circuit}Circuit diagram of the 3500-eV electron gun. Everything inside the red dashed box is contained in two 19-inch racks. The feedback loops (Fig.~\ref{fig:loop}) for the voltages of the cathode and of electrode P1 are not shown. These details are provided in Fig.~\ref{fig:filter}}
		\end{figure}

\begin{figure}[htp]
	\centering\includegraphics[width=1.0\textwidth]{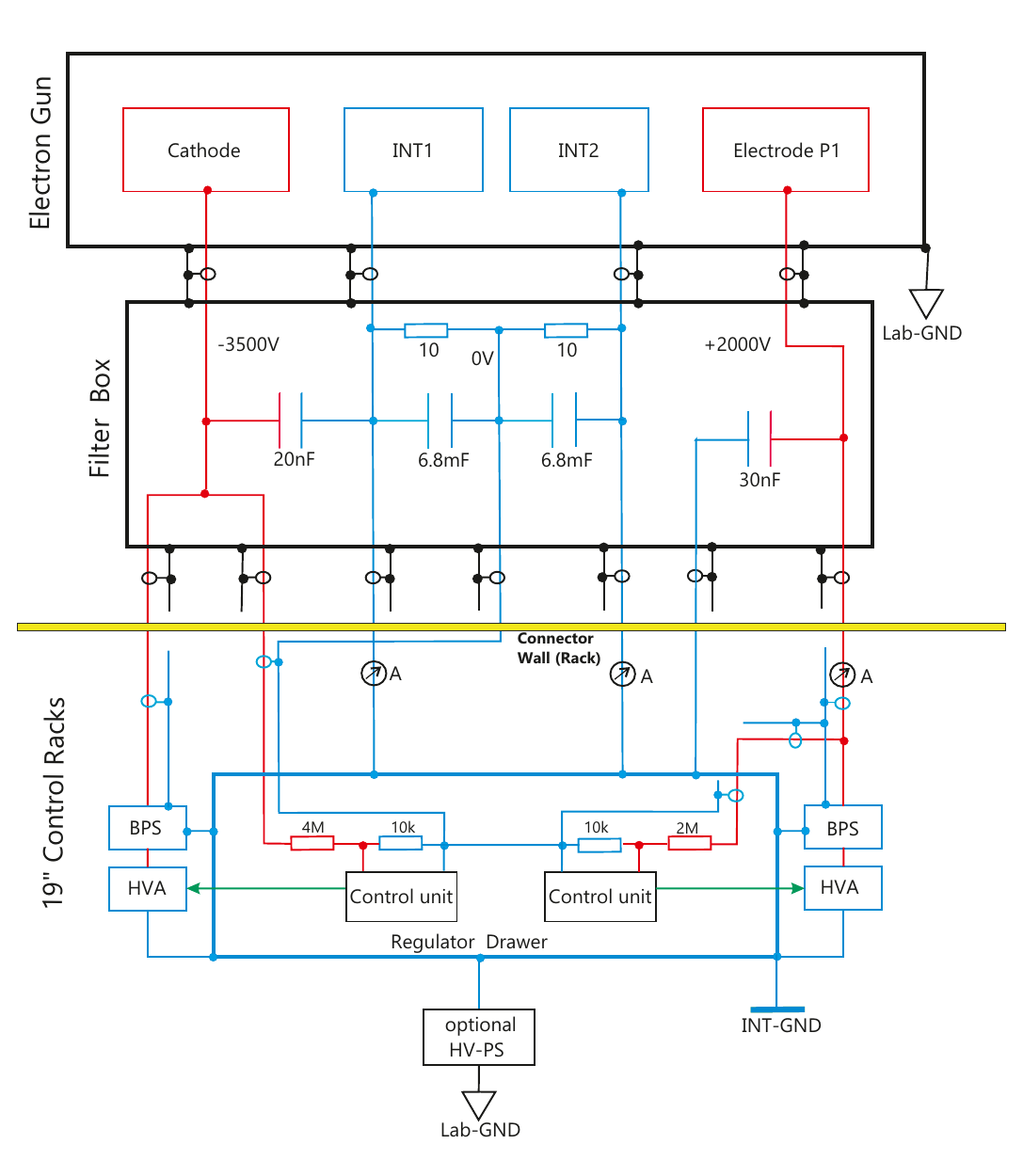}
	\caption{\label{fig:filter}Detailed wiring scheme of cathode, electrode P1, and interaction region showing the respective voltage dividers, the control units of the high-voltage amplifiers (HVA) which are connected in series with a base power supply (BPS). An optional high-voltage power supply (HV-PS) can be used to put the entire arrangement on a potential with respect to the ion beam line  which is an laboratory ground potential (Lab-GND). The filter box is damping the electric noise from the laboratory environment. The time constants are chosen such that the control unit can still act on a sub-millisecond time scale.}
\end{figure}

	\end{appendices}
	\twocolumn
	%%===========================================================================================%%
	%% If you are submitting to one of the Nature Portfolio journals, using the eJP submission   %%
	%% system, please include the references within the manuscript file itself. You may do this  %%
	%% by copying the reference list from your .bbl file, paste it into the main manuscript .tex %%
	%% file, and delete the associated \verb+\bibliography+ commands.                            %%
	%%===========================================================================================%%
	
	%\bibliography{literatur.bib}
	
%% BioMed_Central_Bib_Style_v1.01

\end{document}